\documentclass[sigconf]{acmart}

\AtBeginDocument{%
  \providecommand\BibTeX{{%
    \normalfont B\kern-0.5em{\scshape i\kern-0.25em b}\kern-0.8em\TeX}}}

\copyrightyear{2020}
\acmYear{2020}
\setcopyright{iw3c2w3}
\acmConference[WWW '20]{Proceedings of The Web Conference 2020}{April 20--24, 2020}{Taipei, Taiwan}
\acmBooktitle{Proceedings of The Web Conference 2020 (WWW '20), April 20--24, 2020, Taipei, Taiwan}
\acmPrice{}
\acmDOI{10.1145/3366423.3380242}
\acmISBN{978-1-4503-7023-3/20/04}

\usepackage{multirow}
\usepackage{booktabs}
\usepackage{ifthen}
\usepackage{color}
\usepackage{url}
\usepackage{amsmath}
\usepackage{amssymb}
\usepackage{xspace}
\usepackage[T1]{fontenc}
\usepackage{paralist}
\usepackage{enumerate}
\usepackage[linesnumbered,ruled,vlined]{algorithm2e}
\usepackage{array,multirow,graphicx}
\usepackage{float}
\usepackage{balance}
\usepackage{tikz}
\usepackage{calc}
\usepackage{subfigure}
\usepackage{listings,amsfonts}
\usepackage{amsmath,xcolor,pifont}
\usepackage{url}
\usepackage{xspace}
\usepackage{hyperref,epsfig,endnotes}
\usepackage{xcolor}
\usepackage{array}
\usepackage{paralist}
\usepackage{multirow,makecell}
\usepackage[normalem]{ulem}
\usepackage[misc]{ifsym}

\definecolor{yellow}{RGB}{255,255,153}
\definecolor{grey}{RGB}{224,224,224}

\newboolean{showcomments}
\setboolean{showcomments}{true}
\ifthenelse{\boolean{showcomments}}
 { \newcommand{\mynote}[2]{
      \fbox{\bfseries\sffamily\scriptsize#1}
        {\small$\blacktriangleright$\textsf{\emph{#2}}$\blacktriangleleft$}}}
        { \newcommand{\mynote}[2]{}}

\definecolor{DarkOrange}{rgb}{0.8,0.3,0.0} 
\definecolor{DarkCyan}{rgb}{0.0, 0.55, 0.55}

\newcommand{\tool}[1]{\textit{MadDroid}}

\begin{document}
\title{MadDroid: Characterizing and Detecting Devious Ad Contents for Android Apps}

\author{Tianming Liu}
\authornotemark[1]
\affiliation{%
  \institution{Beijing University of Posts and Telecommunications, China}
}

\author{Haoyu Wang}
\authornote{The names of the first two authors are in alphabetical order. Haoyu Wang is the corresponding author (haoyuwang@bupt.edu.cn).}
\affiliation{%
  \institution{Beijing University of Posts and Telecommunications, China}
}

\author{Li Li}
\affiliation{%
  \institution{Faculty of Information Technology, Monash University, Australia}
}

\author{Xiapu Luo}
\affiliation{%
  \institution{The Hong Kong Polytechnic University, HongKong}
}
  
\author{Feng Dong}
\affiliation{%
  \institution{Shenzhen Institutes of Advanced Technology, CAS, China}
}
  
\author{Yao Guo}
\affiliation{%
  \institution{MOE Key Lab of HCST, Peking University, China}
}

\author{Liu Wang}
\affiliation{%
  \institution{Beijing University of Posts and Telecommunications, China}
}
  
\author{Tegawend{\'e} F. Bissyand{\'e}}
\affiliation{%
  \institution{University of Luxembourg, Luxembourg}
}

\author{Jacques Klein}
\affiliation{%
  \institution{University of Luxembourg, Luxembourg}
}

\renewcommand{\shortauthors}{Liu and Wang, et al.}



\begin{abstract}

Advertisement drives the economy of the mobile app ecosystem.
As a key component in the mobile ad business model, mobile ad content has been overlooked by the research community, which poses a number of threats, e.g., propagating malware and undesirable contents. 
To understand the practice of these devious ad behaviors, we perform a large-scale study on the app contents harvested through automated app testing. In this work, we first provide a comprehensive categorization of devious ad contents, including five kinds of behaviors belonging to two categories: \emph{ad loading content} and \emph{ad clicking content}. 
Then, we propose \tool{}, a framework for automated detection of devious ad contents. \tool{} leverages an automated app testing framework with a sophisticated ad view exploration strategy for effectively collecting ad-related network traffic and subsequently extracting ad contents. We then integrate dedicated approaches into the framework to identify devious ad contents.
We have applied \tool{} to 40,000 Android apps and found that roughly 6\% of apps deliver devious ad contents, e.g., distributing malicious apps that cannot be downloaded via traditional app markets. Experiment results indicate that devious ad contents are prevalent, suggesting that our community should invest more effort into the detection and mitigation of devious ads towards building a trustworthy mobile advertising ecosystem.
\end{abstract}

\maketitle

\section{Introduction}
\label{sec:introduction}
The mobile app ecosystem has seen rapid growth in the past few years.
Google Play and third-party app markets host millions of apps~\cite{AppNumber, wang2018beyond}.
Most apps on markets are free.
Besides, there is also a trend showing that more and more paid apps have been released as free ones by their developers~\cite{wang2019understanding, wang2018android}, suggesting that the business model in free apps offers potentially more attractive revenue. In most cases, while users do not pay to install and run the apps, developers can still monetize through displaying advertisements (or ad in short) on app User Interfaces (UI).
It is estimated that the size of the global mobile ad market would reach 215 billion US dollars by 2021, which will represent 72\% of the total digital budgets~\cite{MobileAdMarket}.

Unfortunately, the mobile ad business model has been abused by malicious individuals to make undue benefits.
For example, unscrupulous app developers are attempting to cheat both advertisers and users with fake or unintentional ad clicks so as to earn profits~\cite{dong2018frauddroid, liu2014decaf, crussell2014madfraud, ClickDroid}. As revealed by a recent report, mobile advertisers have approximately lost 1.3 billion US dollars due to ad fraud in 2015 alone~\cite{MobileAdFraud}, making research on malicious mobile advertisement a critical endeavor for sanitizing app markets~\cite{li2017static, li2017understanding}.

Fortunately, the research community becomes increasingly interested in this area with a variety of research directions targeting the ecosystem of mobile ads. For example, researchers have investigated topics such as automated detection of ad networks~\cite{li2017libd, ma2016libradar, derr:ccs16, li2016investigation}, security and privacy analysis of ad libraries~\cite{pearce2012addroid, grace2012unsafe, derr2017keep}, and detection of mobile ad frauds~\cite{liu2014decaf, ClickDroid, dong2018frauddroid, crussell2014madfraud}. Nevertheless, these studies have so far targeted mobile ad issues from the perspectives of either \emph{app developers} or \emph{ad networks}. The latter plays the role of trusted intermediary platforms for connecting \emph{mobile advertisers} to \emph{app developers} by providing toolkits (e.g., ad SDKs) to be embedded in apps. The perspective of {\em mobile advertisers} themselves, who provide {\em ad contents} and pay ad networks, has been rarely studied. 

Despite being a key component in the mobile ad business model, mobile {\em ad content} has been overlooked by the research community. 
Yet, ad content poses a number of threats. On one hand, ad content downloaded at runtime from trusted ad networks could serve as a channel for attackers to distribute undesirable contents or even malware. For example, even Google Play apps have been reported to display porn ads~\cite{GooglePlayPorn, GooglePlayPorn2}. 
Recent reports also suggested that some ad contents actually come with the CoinMiner malicious script, which uses the device's physical resources in the background to mine digital currency~\cite{coinminerad}. 
On the other hand, besides the ad content itself, some unwanted payload may be triggered when the user interacts with the ad content. For instance, the ad clicking event could redirect the current execution page to a malicious website. Overall, we refer to such ad contents as {\bf \em devious}, since they are deceitful for all parties (i.e., for app users, for app developers, and potentially for ad networks when they are unaware of this bad behavior of mobile advertisers).

To the best of our knowledge, there lacks an in-depth study on both \emph{ad loading contents} and \emph{ad clicking contents}. 
The closest studies including Chen \textit{et al.}~\cite{chen2019revisiting} and Shao \textit{et al.}~\cite{shao2018understanding} only examine ad clicking contents, and thus have several limitations (detailed in the evaluation section) and overlook numerous ad clicking contents. 
In this paper, we fill this gap by performing a comprehensive study of mobile ad contents, aiming to understand the state of practice in devious ad contents and devise practical techniques for preventing their spread in the mobile ecosystem.
To this end, we first present a systematic approach to categorizing devious mobile ad contents based on a thorough investigation of ad-related policies and reports (\textbf{Section~\ref{sec:motivationandcategory}}). 
Then we design and implement \tool{}, a prototype framework for automated detection of devious mobile ad contents (\textbf{Section~\ref{sec:approach}}).
\tool{} leverages a dedicated automated app testing approach to explore ad views in an app, based on a sophisticated ad-first strategy (\textbf{Section~\ref{sec:TCM}}).
While exploring mobile ads, \tool{} records any network traffic and collects contents exchanged between mobile ad networks, advertisers and user devices. By hooking the HTTP-related APIs in the Android framework, \tool{} manages to precisely locate ad traffic from all recorded traffic (\textbf{Section~\ref{subsec:cem}}). 
Finally, we implement a number of specialized approaches (\textbf{Section~\ref{sec:ADD}}) to detect devious ad contents.

To summarize, we make the following main contributions:

\begin{itemize}
\item \textbf{A novel ad traffic identification approach}. We present an HTTP hooking approach to \emph{iteratively} build a mapping between ad libraries and ad hosts.
This mapping enables our approach to precisely identify ad traffic from general network traffic. Experimental results suggest that, our approach outperforms state-of-the-art ad traffic identification methods significantly, i.e., we have identified three times of the ad hosts and increased the collection of ad contents by 126\%.

\item \textbf{A comprehensive detection framework.} We propose \tool{}, a framework to detect devious mobile ad contents. To the best of our knowledge, this is the first attempt in the literature to detect five groups of devious mobile ad contents.

\item \textbf{A large-scale study in the wild.} We conduct a large-scale empirical evaluation on the usefulness and effectiveness of \tool{}. By applying \tool{} to 40,000 apps, we find roughly 6\% of apps (2,322) that deliver devious ad contents.
We have released the dataset and experiment results to the research community at:

\begin{center}
\url{https://github.com/MadDroid-2020/MadDroid-WWW} 
\end{center}

\end{itemize}

\section{Background and Terminology}
		\vspace{-0.1in}

\begin{figure}[!ht]
	\centering
	\includegraphics[width=0.6\linewidth]{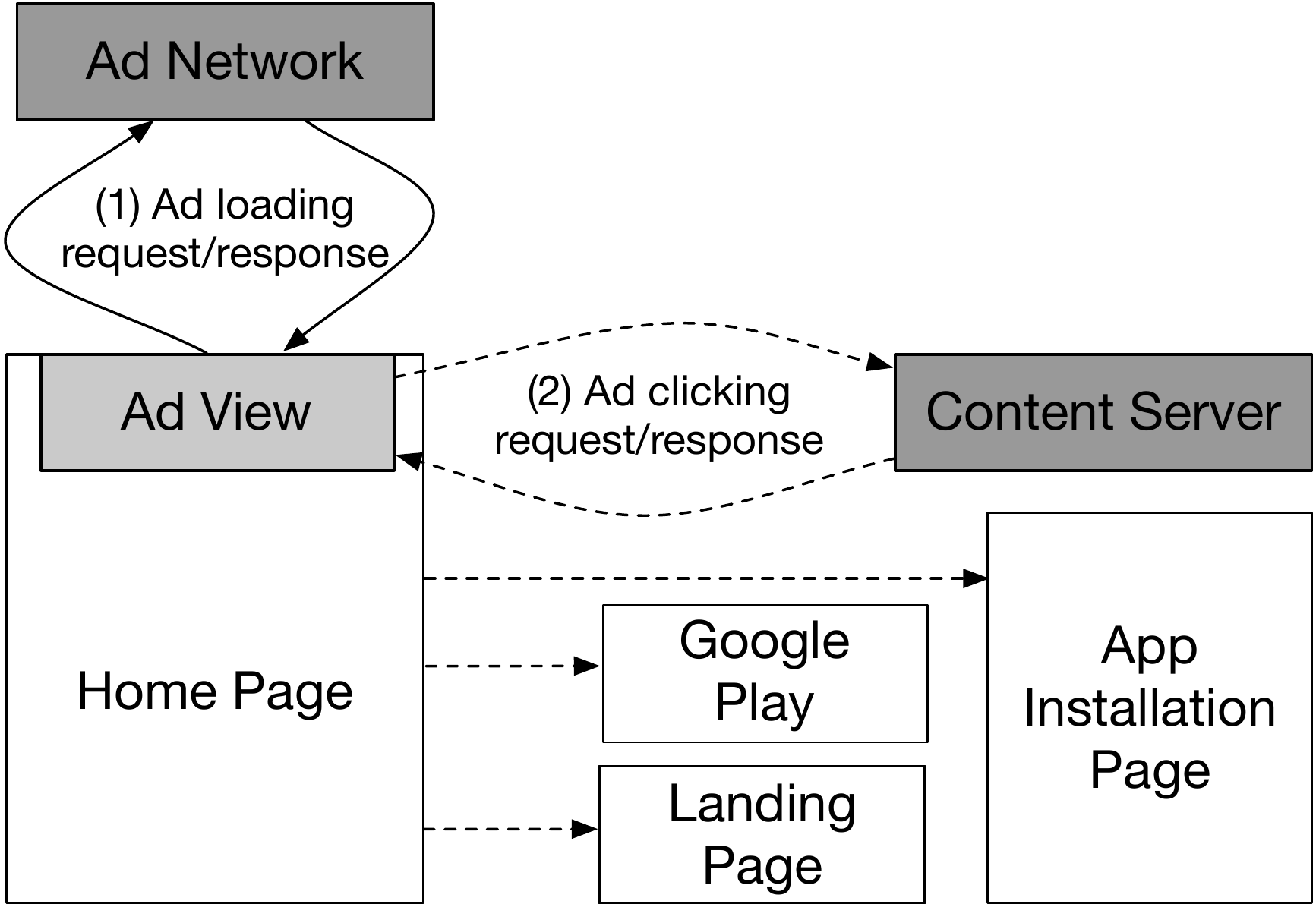}
		\vspace{-0.1in}
	\caption{The general working process of mobile ads. }
	\vspace{-0.15in}
    \label{fig:term}
\end{figure}

In order to clarify the meaning of specific terms used in this paper, and to help readers get an overall understanding of how mobile ads work, we briefly describe the workflow of mobile ad delivery on users' device interfaces. 
Figure~\ref{fig:term} illustrates the overall workflow. 
For simplicity, we will refer to any graphical user interface where an ad can be displayed as a \textbf{Home Page}. 
When such a page appears on the foreground of a device's screen (e.g., after a menu item is selected), an ad-related HTTP request is sent in an attempt to fetch {\em ad content} from an \textbf{Ad Network}. In the mobile ecosystem, the Ad Network plays the role of a trusted intermediary platform that connects \textbf{advertisers} to \textbf{app developers} by providing ad libraries (e.g., Google's AdMob) to be embedded in app code for fetching and displaying ads at runtime.
In response to the ad-related HTTP request, the Ad Network may serve for example an image that will be used on the Home Page to update an ad view.

Once the ad view is displayed on the Home Page, users can click it to observe its content.
Normally, when the ad is clicked, it will again trigger another ad-related HTTP request that attempts to fetch additional {\em ad contents} from a \textbf{Content Server}, which may be hosted by advertisers or other third-parties.
There are three types of ad contents that Content Servers recurrently push to users:
\begin{enumerate}
\item A redirection link that switches the current Home Page to a so-called \textbf{Landing Page} for displaying the ad information, such as an online shopping page where the user can purchase the items that were usually advertised on the Home Page. 
\item  A deep-link that switches the current Home Page to \textbf{Google Play} for helping users install advertised apps.
\item Automatic download of a file. Typically, this is an APK file. When the APK downloading is completed, the current Home Page is switched to an \textbf{App Installation Page}, where users can decide whether to install the downloaded app.
\end{enumerate}

As shown in Figure~\ref{fig:term}, there are two types of ad-related HTTP requests: one as {\em Ad loading} request and the other as {\em Ad clicking} request. Unfortunately, the ad content served in response to both requests may be comprised of devious artifacts that may threaten the security and privacy of app users.

\section{Motivation and Categorization}
\label{sec:motivationandcategory}
We first describe a real-world example of devious ad contents that we have encountered on a popular racing game app. 
We then create a categorization of devious mobile ad contents based on a thorough investigation of ad-related policies and reports, which will drive the implementation of techniques for identifying devious ad contents.

\subsection{Motivating Example}
\label{subsec:motivation}

\begin{figure}[!ht]
	\centering
	\includegraphics[width=0.4\linewidth]{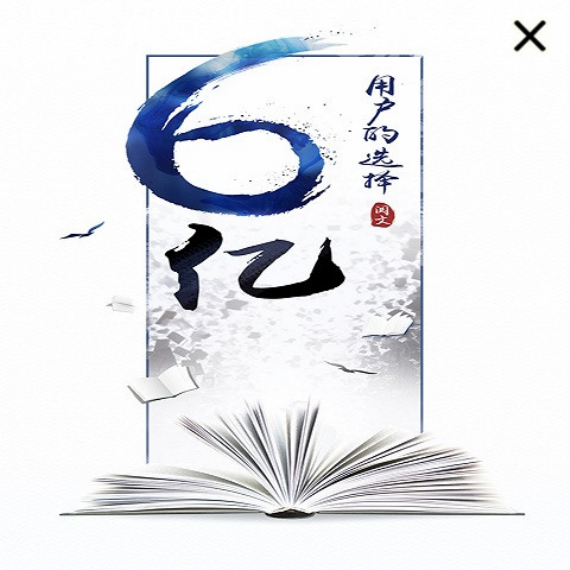}
  \vspace{-0.2in}
	\caption{An example of click-deceptive Image.}
      \vspace{-0.15in}
    \label{fig:motivation}
\end{figure}

While using the free racing game app \emph{Speed Racing Ultimate}~\cite{motivatingVT}, it is not uncommon to see ads appearing on the foreground. 
Figure~\ref{fig:motivation} provides the screenshot of an example ad view observed by one of the authors while playing the game. 
 At the top right corner, there is a {\em cross} symbol ({\large $\times$}), which conventionally suggests that the ad view can be closed by clicking at this location.
 Once clicked, however, a redirection is triggered and the current home page is replaced by a landing page where ad content is displayed.
 At first, one may suspect that the user failed to properly click on the correct location, instead clicked on the actual ad, justifying the behavior. Nevertheless, after several failed attempts, the user concludes that the ``close'' functionality is not supported, or at least not working as expected, via the cross symbol. Further manual investigations into the ad later revealed that \emph{the ({\large $\times$}) symbol is actually embedded in the image}. This demonstrates a deceitful behavior as the purpose of the cross symbol was never to close the ad but to trick users into clicking on the ad. Such devious ads are increasingly frequent in practice, however, studies about them are scarce in the literature.
 
\subsection{Categorization of Devious Ad Contents}

\begin{figure}[!ht]
	\centering
	\includegraphics[width=0.95\linewidth]{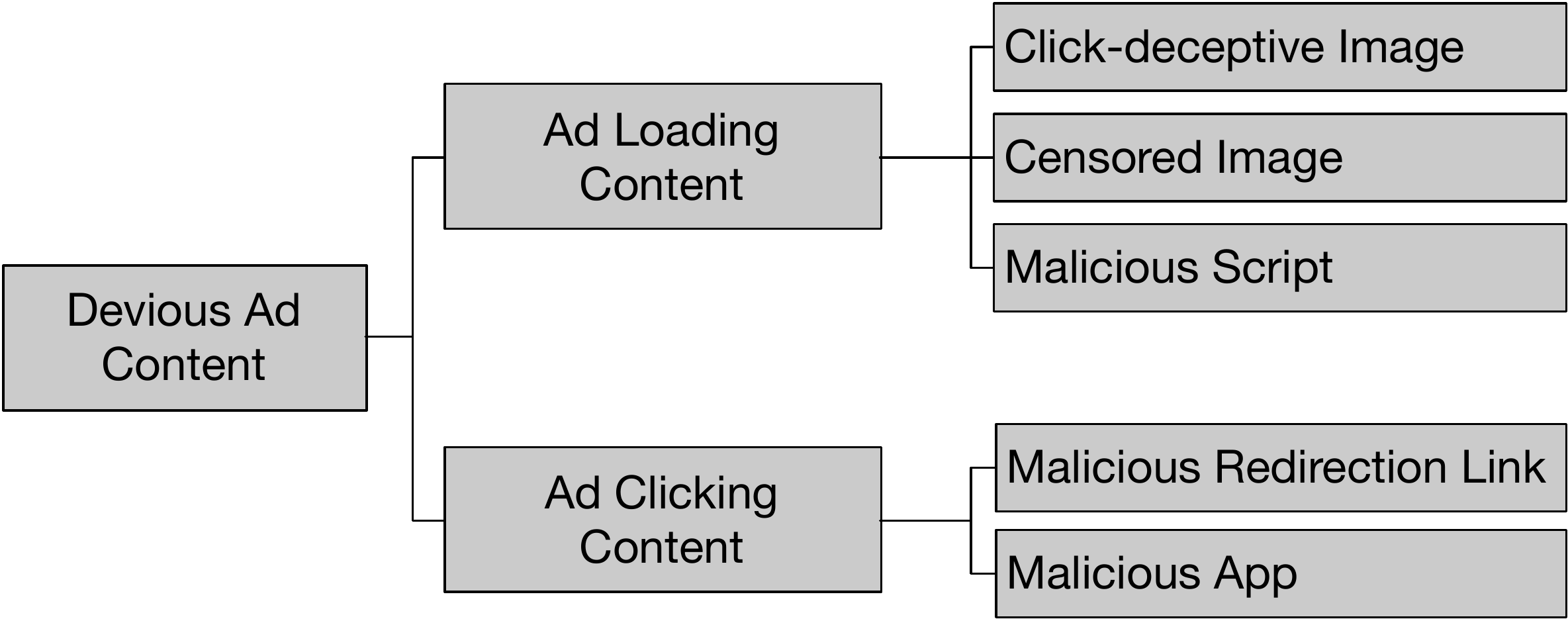}
	\caption{The five categorized groups of devious ad contents.}
	\vspace{-0.1in}
    \label{fig:taxonomy}
\end{figure}

The consequence of the redirection triggered by the devious ad content example presented above was a simple annoyance for users. 
However, we can imagine scenarios where such a redirection lands on malicious payload being performed. 
Thus, motivated by such possibilities, we decided to conduct a systematic study of the current devious mobile ad contents.
To categorize such contents, we first investigate the undesirable mobile ad contents from: (1) the policies related to mobile ad contents of popular app markets~\cite{googleplaypolicy, TencentDeveloper, Huawei, MedicalGoogle, AliDeveloper}, (2) media reports in news outlets~\cite{drivebydownload, GooglePlayPorn,GooglePlayPorn2, coinminerad, Malvertising, malvertising2, spammyads}, and (3) some real-world apps that host devious ad contents. Based on our empirical investigation, we summarize the observed devious ad contents into five (5) groups enumerated in Figure~\ref{fig:taxonomy}.

Note that three groups, namely {\em Click-deceptive Image},  {\em Censored Image}, and {\em Malicious Script}, are related to ad contents obtained following the Ad Loading request, while the remaining two, namely {\em Malicious Redirection Link} and {\em Malicious App} are related to ad contents obtained after an Ad Clicking request. 

\textbf{(1) Click-deceptive Image:}
As shown in Section~\ref{subsec:motivation}, devious ad networks (or advertisers) may provide, as ad content, a click-deceptive image, where a ``cross'' symbol ({\large $\times$}, or similar images) is directly embedded in the ad image aiming at tricking users into clicking on it to close the ad view. Normally, the ``close'' button of an ad is displayed as a separate image: on one hand, this allows ad networks (or app developers) to set events (such as click to close) that are independent from events associated to the ad image (such as click to follow a link); on the other hand, setting a separate image offers the opportunity to display the ``close'' button after a delay of several seconds, giving enough time for users to notice the ad.

\textbf{(2) Censored Image:} 
Censored Image refers to such ad images that fall under censorship with respect to state legislation or the market policy. In this work, we enumerate the cases of \emph{Gambling}, \emph{Violence}, \emph{Medical}, and \emph{Pornographic} images, which might be prohibited. Google itself explicitly warns developers that gambling advertising should abide by local gambling laws and industry standards~\cite{googleplaypolicy}. Similarly, Google also disallows the presentation of violent ads as they are not appropriate for children, while some medical-related contents cannot be advertised at all~\cite{MedicalGoogle}.
Adult ads also need to comply with certain policies: for example, it is not allowed to distribute ad contents that may be interpreted as promoting a sexual act in exchange for compensation in many countries. 
Besides Google Play, many third-party app markets~\cite{AliDeveloper, Huawei, TencentDeveloper} do not allow advertising of Gambling and Pornographic contents.

\textbf{(3) Malicious Script:}
Mobile ads, which are usually displayed via the WebView widget in Android, can legitimately run code to interact with the host app. For example, a code fragment can be included to remove the ad after the close button is clicked.
Unfortunately, devious ad networks may inject malicious scripts in the ad. For example, the 360 Fenghuo Lab has reported that some ad networks distribute devious ad contents through which they mine bitcoins on users' devices, without their knowledge~\cite{coinminerad}. 

\textbf{(4) Malicious Redirection Link:}
Some mobile ads, after clicked, may jump to landing pages where malicious contents are presented to the users. When such redirected content is clicked, the security and privacy of the user may be in jeopardy.

\textbf{(5) Malicious App:}
This group refers to such mobile ads that, when clicked, may download malicious Android apps into the user device. In this scenario, devious ad contents appear as an attractive means to distribute malware on user devices. 

\vspace{-0.1in}
\subsection{Challenges}

In this work, we aim at proposing an effective approach to detect these aforementioned types of devious ad contents from Android apps.
It is nevertheless non-trivial to achieve this automatically.
There are at least three challenges that need to be effectively addressed.
The three challenges are summarized as follows.

\textbf{How to automatically trigger and collect ad content?} Mobile ad contents could be collected at the time when the ad is fully loaded or consumed, which requires not only triggering the appearance of mobile ads but also clicking the presented ads.
Unfortunately, mobile ads could be delivered in different sizes (e.g. Banners, Interstitials, Full Screens), different carrier widgets (e.g. WebView, ImageView, ViewFlipper), different numbers and places (one or multiple, within the same UI state or different states), sophisticated approaches hence are needed to effectively traverse ads in apps while ensuring good coverage.

\textbf{How to efficiently pick out ad traffic from general network traffic?}
Mobile ad contents can be extracted from the network traffic, specifically the ad-related traffic (or ad traffic in short).
However, when collecting ad traffic at runtime, general network traffic would be also collected, i.e., non-ad traffic and ad traffic are inevitably mixed. 
Hence, there is a need to design effective approaches to separate ad traffic from the general ones.

\textbf{How to precisely differentiate devious ad contents from normal ad contents?}
With a systematic approach, we have identified and categorized five groups of devious mobile ad contents, which respectively need specialized techniques to characterize.
Considering that new groups of devious ad contents can be added in the future, the detection approach should not only be inclusive (e.g., cover all the devious groups), but also extensible (e.g., can be easily extended to cover new devious groups).

\section{Approach}
\label{sec:approach}

\begin{figure}[t]
	\centering
	\includegraphics[width=\linewidth]{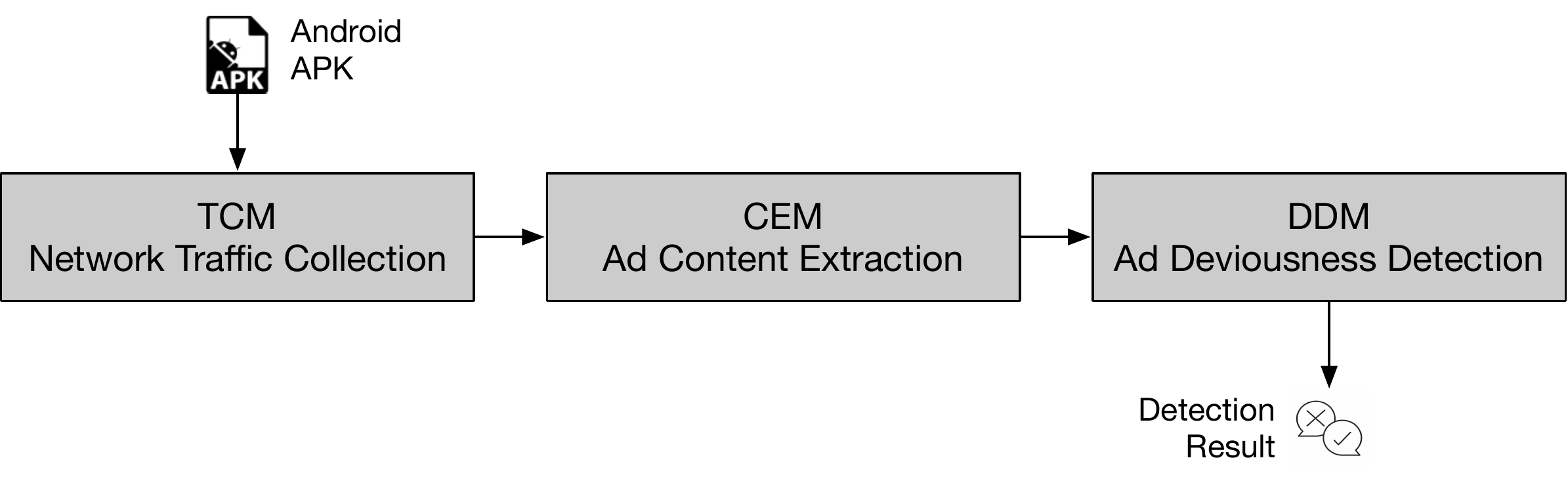}
	\vspace{-0.2in}
	\caption{Overview of the \tool{} framework.}
	\vspace{-0.2in}
    \label{fig:overview}
\end{figure}

Figure~\ref{fig:overview} depicts the essential modules of the workflow in our proposed \tool{} framework. Towards detecting devious ad contents that are delivered to an input Android app, we propose an architecture with three modules that respectively address the aforementioned three challenges:

\begin{itemize}
	\item {\bf TCM}:  a network Traffic Collection Module, which focuses on traffic generated as part of the ad loading or interaction phases. This module requires careful design as it requires dynamic execution which, in order to be effective, must be focused on covering mainly ad-involved UIs.
	\item {\bf CEM}: an ad Content Extraction Module, which learns to identify, among exchanged traffic, which ones are about loading content that must be extracted. It further explores contents that are delivered after an ad view is clicked.
	\item {\bf DDM}: an ad Deviousness Detection Module, which finally analyzes the extracted ad contents to identify devious ones. Given the diversity of devious ad contents, this module implements specialized detection schemes with adapted techniques ranging from character recognition to deep learning. 
\end{itemize} 

This modular architecture enables flexibility for extension and maintenance. Given that the categorization presented in this paper is based on the currently known devious ad contents, when other devious ad contents and distribution scenarios are uncovered, each module can be appropriately extended to take them into account. In the remainder of the section, we describe in detail our approach for implementing each module.

\subsection{Network Traffic Collection}
\label{sec:TCM}

The TCM module implements the first step in the \tool{} framework. Its objective is to harvest all the network traffic that is involved in operations for delivering ads on a given app. 
This traffic carries not only data from exchanges between ad networks and the home page view (i.e., when the app is being loaded), but also data from exchanges between the home page and the advertiser's content server (i.e., when the user interacts with the ad). Thus, given an Android apk file, TCM must visit all app UI pages where ads are likely to be loaded, and then explore an interaction with such ads to collect data in the reached landing page.

For scalability reasons, TCM must implement an effective and automated strategy for covering all ad views in an app. Nevertheless, although it is labor-intensive and time-consuming to implement the exploration manually, it is also non-trivial to achieve automation via traditional automated app testing~\cite{kong2018automated}. 
Indeed, state-of-the-art approaches in Android, such as MonkeyRunner~\cite{Monkeyrunner}, generate random test cases that are not ad-specific: the majority of dynamic execution scenarios will then be wasted for exploring irrelevant UI states. 
As empirically demonstrated by Suman Nath~\cite{nath2015madscope} on a set of ad-supported apps, over 90\% of the automatically explored UI states are not ad-involved pages.

To overcome the efficiency challenge in rapidly and quasi-exclusively focusing on relevant UI states, we propose to tune the exploration strategy by generating ad-intensive test cases, i.e., by favoring ad views. 
We refer to it as an {\em ad-first exploration} strategy. We build on the finding of a recent study~\cite{nath2015madscope} that most ads are displayed in the main UI page and on the exit UI page. 
Our ad-first exploration strategy thus attempts to prioritize the views of these pages, and further rely on a breadth-first search algorithm where the views in a page are reordered, i.e., ad views are prioritized.

\begin{figure}[t]
	\centering
	\includegraphics[width=0.95\linewidth]{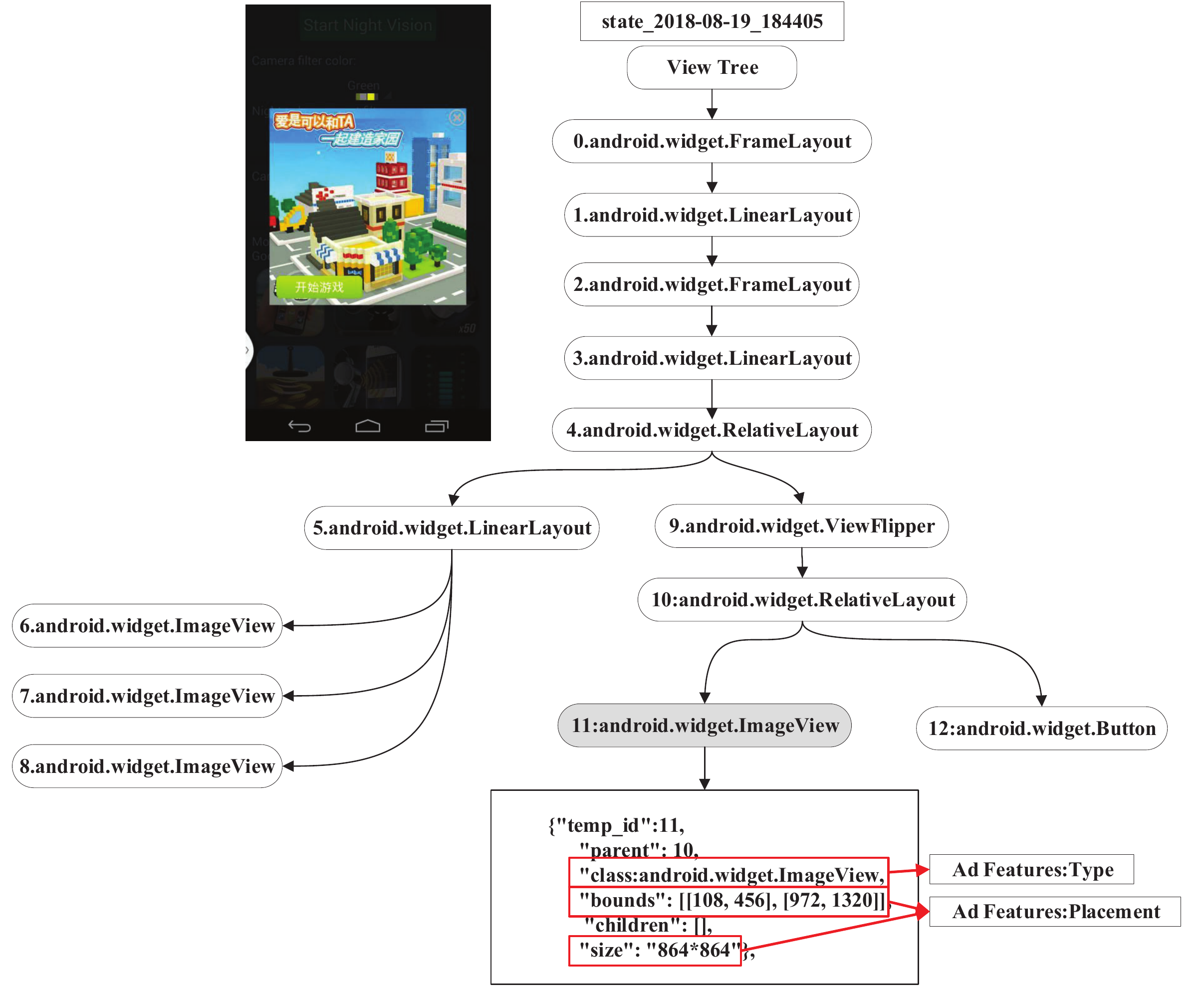}
	\vspace{-0.1in}
    \caption{An example of a view tree.}
    \vspace{-0.2in}
    \label{fig:viewtree}
\end{figure}

Views are identified by traversing the nodes in a view tree that can be obtained from a given UI state (i.e., a GUI page at a given time in app execution).
Figure~\ref{fig:viewtree} shows an example of a view tree. 
The root node represents the base {\em layout view} on top of which upper views are placed. Parent nodes are containers to child nodes that are subject to users' manipulations.
Each node is tagged with basic view information such as position, size, class name, etc. Inspired by a most recent work~\cite{dong2018frauddroid}, we use such attributes to identify which nodes among the leaf nodes are likely to be ad view nodes. Specifically, during our exploration, to ensure good coverage, we prioritize and click on each view that falls in the class of WebView, ImageView or ViewFlipper.

Based on the results of the ad-first exploration, dynamic execution of ad-related UI states will lead to a large collection of network traffic.
Unfortunately, at this stage, the collected traffic contains not only ad-related traffic but also non-ad related ones (such as data exchanged for app analytics).
There is hence a strong need to precisely distinguish between ad and non-ad traffic, in order to correctly extract ad content.
To this end, we propose a framework runtime hooking approach to achieve this purpose. Details will be given in the next subsection.

\subsection{Ad Content Extraction}
\label{subsec:cem}

The CEM module analyzes the traffic collected through TCM in order to extract relevant content for further assessment. Indeed, by default, TCM collects any traffic that occurs while the ad is being loaded or after it is clicked. 
Since we are interested in traffic carrying ad contents, CEM must dismiss all traffic that is not related to advertisements (e.g., parallel traffic from core app functionality).
To this end, the first step taken in CEM is to identify ad-related HTTP requests/responses, considering those that are done as part of exchanges with ad networks.
Take Table~\ref{tab:listing} as an example, a simplified list of HTTP requests harvested from the execution of app \emph{com.bbsoft.InternetPolyglot} is illustrated and \emph{startappservice.com} is known as an ad-domain.
The first step is hence to highlight such ad-domain related HTTP requests (cf. lines 1, 3 and 4).

\begin{table}[!h]
\caption{Simplified list of harvested requests (domain + path)}
\centering
\resizebox{0.9\linewidth}{!}
{\tt
\begin{tabular}{l|ll}

1&  AD-DOMAIN  & 	info.static.startappservice.com  \\
&    & ~~~~~~~~~~~~~~~~~~	/1.4/getadsmetadata \\
2&  NON-AD  & 	data.flurry.com	/aap.do \\
 & //ad-load &  \\
3& AD-DOMAIN  &	req.startappservice.com	/1.4/gethtmlad \\
4& AD-DOMAIN &	imp.startappservice.com	 \\
&  &	~~~~~~~~~~~~~~~~~~	/tracking/adImpression \\
 & //ad-click &  \\
5& NON-AD &	cl.untildogtop.com	/t/clk \\
6& NON-AD &	my1trk.com	/redirect/action \\
&  &	~	/1InYjNywuJnNnYTwiKHNmf3BlZ2E\_eQ\_Pyi \\
7& NON-AD &  	www.spyoff.com	/geo \\
\end{tabular}
}
\vspace{-0.1in}
\label{tab:listing}
\end{table}

\begin{figure}[!ht]
	\centering
	\includegraphics[width=\linewidth]{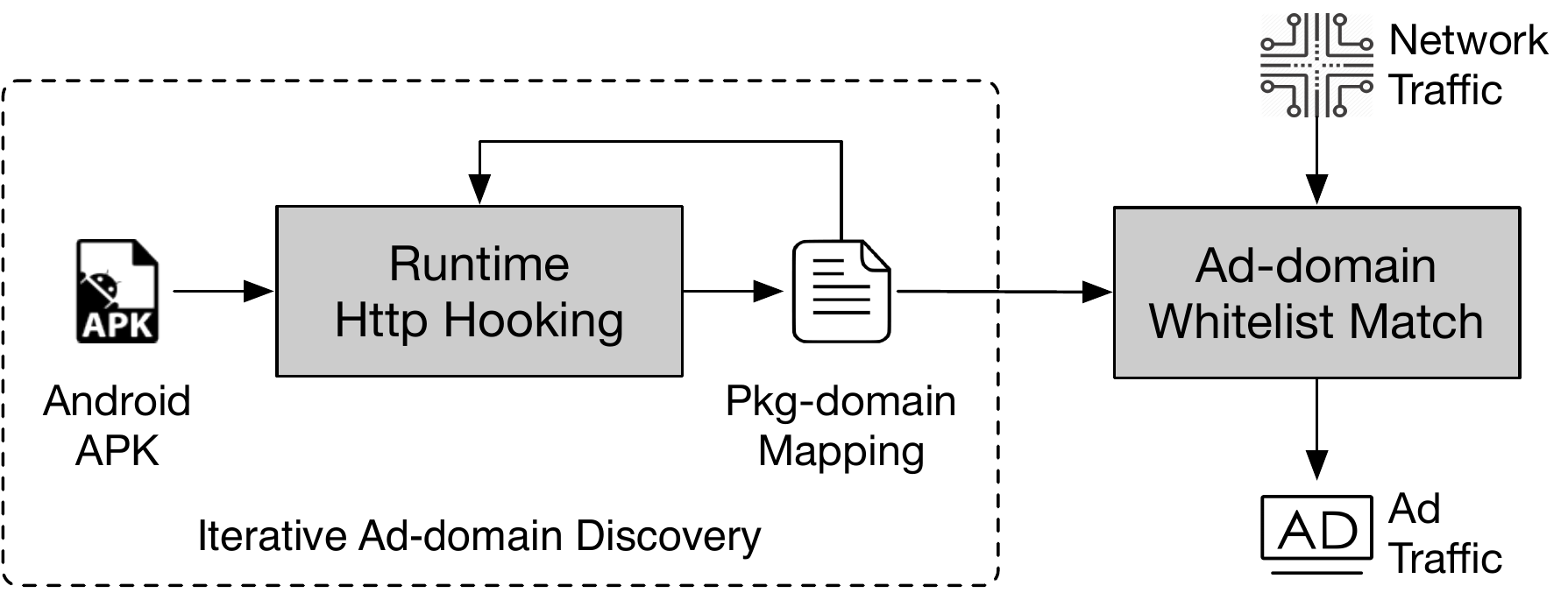}
	\vspace{-0.1in}
	\caption{Runtime hooking approach for locating ad traffic.}
	\vspace{-0.1in}
    \label{fig:hook}
\end{figure}

As shown in Figure~\ref{fig:hook}, given an ad-domain whitelist, it would be straightforward to pick out ad-load traffic from a collection of network traffic.
Unfortunately, it is not easy to manually build such a whitelist of ad domains, mainly due to two reasons. On one hand, there are a plethora of ad libraries and new ad networks might continuously join the ecosystem. On the other hand, we empirically found that, for a given ad library, the domain names of ad networks may change, and even one ad library may correspond with a number of domain names, making it hard to label a complete and accurate list of ad-domain names. For example, we found that the ad network ``daoyoudao''~\cite{Daoyoudao} has dozens of ad-domain names, including ``daoudao.com'',  ``guiji.com'', ``133155.com'', ``161161.com'' and ``150155.com'', etc.

Therefore, we propose to develop in CEM a runtime HTTP hooking approach (as shown in Figure~\ref{fig:hook}) for iteratively identifying ad relevant domain names, so as to locate ad-relevant traffic.
Our approach dynamically hooks all the HTTP-related methods at the framework level. Following the same ad-first exploration approach detailed in the previous section, when an HTTP-related method is reached, the hooking module will record the current execution stack trace and the URL associated with the HTTP method.
Following the dumped stack trace, our approach can automatically locate the package that initiates the HTTP connection and build a mapping (hereinafter referred to as pkg-domain mapping) from packages to URL domains.
If the package belongs to a known ad library, all the domains triggered by this package will be regarded as ad-domains and recorded into the mapping.
Similarly, if the domain matches one of the ad-domains recorded in the mapping, the corresponding package will be flagged as an ad library and hence recorded into the mapping.
By doing so, the runtime hooking approach enables our approach to iteratively grow the whitelist of ad-domains.

Once ad traffic is located from the network traffic collected by TCM, it can unfold the next step of extracting ad contents from all relevant ad-response messages (i.e., those sharing the same session id as the identified ad request messages). Ad contents that are extracted include images and executable scripts.
Such ad traffic that is only relevant to simple message exchange (e.g., sending a message to confirm an ad impression) without carrying actual ad content will be ignored (e.g., lines 1 and 4 in Table~\ref{tab:listing} will be ignored).

With regard to ad click, the response message of an ad-loading request generally includes a URL indicating the target address when the ad is clicked. Indeed, ad clicking requests are supposed to be redirected to the appropriate content server whose domain address is stored in the ad content. More specifically, the domain address is bound to ad click events.
Let us take Listing~\ref{tab:listing} again as an example, after the ad is clicked (line 5), an ad clicking request will be sent to a content server, which returns a redirection link (line 6) that eventually leads to the ad landing page (line 7), which is a VPN app website.
By analyzing the binding information, we can retrieve this address and subsequently identify ad clicking-related traffic.
By dynamically exploring the ad clicking events on installed apps, we can further collect three types of ad contents: (1) \emph{redirection links}: the URL bound to the ad click event might not be the final destination: i.e., the landing page may be reached after several redirections.
(2) \emph{downloaded APKs}: the ad clicking request will trigger a downloading process of non-requested apps.
(3) \emph{Google Play pages}: the ad click will be directed to Google Play to promote the advertised app.

There are at least two means to explore the ad clicking events: (1) by simulating the clicking request (e.g., record the request URLs and then send requests using a browser later) or (2) by actually clicking the ad.
The latter approach is adopted in this work as we have experimentally found that the former approach is likely leading to failures of requests.
For example, we have empirically observed that some redirection links are time-sensitive. The emulated request after a certain time period will simply result in an invalid request.

\subsection{Ad Deviousness Detection}
\label{sec:ADD}

The DDM module in \tool{} is the core component in charge of implementing analysis procedures for assessing the variety of artifacts collected by CEM in order to check against the presence of any devious ad content.
Given that each group of devious ad content presents specific characteristics and detection challenges that require specific detection schemes, we design DDM with a plugin-based system architecture. 
This offers the flexibility to address newly appearing groups of ad contents by integrating an independent plugin implementing the required analysis of ad contents using specialized state-of-the-art techniques.

In the current version of \tool{}, we have already proposed prototype plugins that cover the devious ad content groups. We now detail, for each plugin, the detection strategy that was applied as well as some implementation details.

\subsubsection{\textbf{Click-deceptive Image}}
\begin{figure}[t]
	\centering
	\includegraphics[width=\linewidth]{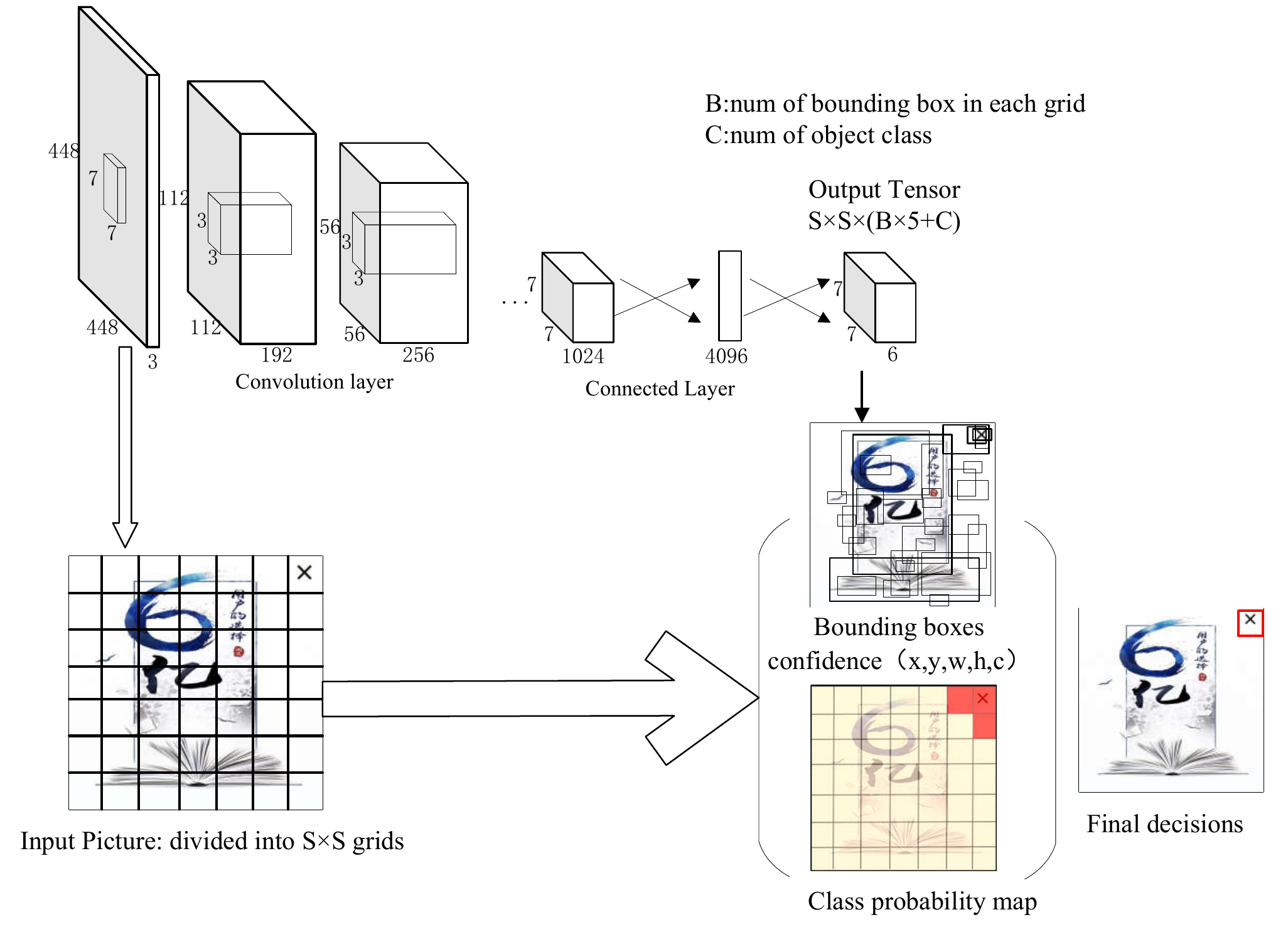}
    \vspace{-0.3in}
    \caption{Click-deceiving picture detection based on YOLO.}
    \vspace{-0.2in}
    \label{fig:yolo}
\end{figure}

The main idea
is to check whether the image actually embeds a ``cross'' symbol. This refers to the problem of recognizing objects in images. Traditional object detection algorithms have shown to be effective for object recognition~\cite{surf, YOLO, yolopaper}. In this work, we adopt the YOLO (You Only Look Once) approach, which is proven to have achieved higher efficiency and accuracy than other approaches~\cite{YOLO, yolopaper}. The work-process is illustrated in Figure~\ref{fig:yolo}. First, the algorithm splits the image into an $S*S$ grid. 
Then, for each grid cell, it predicts $B$ bounding boxes to mark the object, and the confidence for each box.
These predictions are encoded in a tensor of $S \times S \times (B \times 5+C)$ dimension, where $C$ is the number of objects to be recognized. We set $C=1$ in our work as we aim to recognize only a single object.
Finally, YOLO uses a non-maximal suppression approach to choose the box that yields the best prediction score. For more details on the inner-working of the algorithm, we refer the reader to the description in ~\cite{YOLO, yolopaper, yolov3}.

Although object recognition techniques have been proposed for decades in various applications, including face detection, the literature, to the best of our knowledge, does not report any work related to the case of ``cross'' ({\large $\times$}) symbol, a simple but pervasive object.
As a result, there is no public dataset that we can leverage to train our model for the detection of ad click-deceptive Images.
As part of the \tool{} effort, we propose to construct such a training set from scratch. Although we had already harvested some sample images during our manual investigations for the purpose of characterizing devious ad contents, the obtained set is not representative. 
Since normal ads also contain ``cross'' symbol and it is difficult for users to distinguish if the ``cross'' symbol is displayed from an independent image, we collect images from found ads and artificially embed ``cross'' symbols into them.
Note that we have collected more than 100 different kinds of ``cross'' symbol images from normal ads, and we artificially embedded them into images with random positions and random size (within a normal size range). 
Eventually, we obtain a set of 2,375 pictures and record the ground truth (i.e., the actual position of the ``cross'' symbol) via a common object recognition format \texttt{PASCAL Visual Object Classes}~\cite{pascal}.

\subsubsection{\textbf{Censored Image}}
We treat separately the cases of gambling and pornographic/violence/medical devious ad content which are all considered as censored images.

\paragraph{\textbf{Pornographic/Violence/Medical Ad Picture.}}
Image detection has been a hot topic in the research community for decades.
With the recent advances in CV and deep learning, the research line has matured, and many highly effective approaches~\cite{lin2003pornography, marcial2011detection, ulges2011automatic, nsfw} are available.
Given as an input an ad image, Google Vision API~\cite{GoogleVision} will output a range from 1 to 5 (i.e., from very unlikely to likely and very likely) indicating the likelihood of being the image targeted by the analysis (e.g., pornographic, violence, or medical).
In this work, we consider that a given image is a censored one as long as the prediction result is equal or higher than 4, indicating the image is likely or very likely to be a pornographic/violence/medical image.

\paragraph{\textbf{Gambling Ad Picture.}}
Because of the heterogeneity in gambling (e.g., blackjack, poker, etc.), 
it is hard to build a graphical model that captures the ``gambling'' instances accurately. Thus, instead of detecting gambling images graphically, we adopt a simple approach that focuses on the text embedded in the ad images.
To that end, we rely on Optical Character Recognition (OCR)~\cite{OCR} techniques to extract any text from a given ad image and match them against a predefined set of gambling keywords. We consider that an ad is about delivering gambling contents if any of its embedded text tokens match any of the keywords enumerated currently in the prototype plugin implementation. 
The gambling-related keywords are collected from various sources, including the top searched Casino keywords in Google~\cite{CasinoWords}, and the frequently presented words on several online gambling websites (in both English and Chinese). Since similar ad contents displayed on online gambling websites might be also used in mobile apps, when counting the recurrently presented words, we also take into account the words presented in pictures of those websites. Eventually, our gambling-related keyword set contains 100 words in Chinese and English\footnote{Example keywords include gambling, casino, Macau dealer, beauty Croupier, lottery, GoldenFlower (ZHAJINHUA in Chinese), etc.}.

\subsubsection{\textbf{Other Devious Ad Content Groups.}}

Devious ad content for other groups, namely {\em Malicious Script}, {\em Malicious Redirection Link} and {\em Malicious App}, which, contrary to images, have been well investigated in the security community. Hence, given that we build a framework, for our plugin prototypes, instead of reinventing the wheels, we leverage state-of-the-art techniques to detect issues with such artifacts. Specifically, we rely on anti-virus scanners to flag malicious artifact. Concretely, DDM sends these non-image artifacts to VirusTotal~\cite{virustotal}, a free online service that integrates over 60 anti-virus engines, has been widely adopted by the research community~\cite{chen2016following, hu2019dating, suarez2018eight, hu2019want}. Our prototype plugins implement detection schemes where, for each artifact that is sent to VirusTotal will be considered as ad devious content whenever at least three (3) anti-virus scanners flag it as suspicious.

\subsection{Implementation}

The core of the \tool{} framework is implemented in Python. It includes the architectural foundation for gluing the input and output formats of the different modules, as well as for reporting decisions. 
We have implemented a lightweight UI-guided test input generator to dynamically explore Android apps, with a special focus on ad views during exploration, including the enforced ad-first exploration strategy. 
The network traffic is harvested through Fiddler~\cite{fiddler}, which serves as a Man-in-The-Middle (MiTM) service between test devices and the server to decrypt and record all HTTP(S) traffics, which are further sent to customized Fiddlerscripts~\cite{fiddlerscript} to extract specific content from the traffic. 
The runtime hooking is based on the Xposed framework~\cite{Xposed}, which can collect the runtime information of tested Android apps.

\section{Evaluation}

Our evaluation is driven by the following research questions (RQs). 

\noindent
\textbf{RQ1:} Can \tool{} detect devious mobile ad contents?

\noindent
\textbf{RQ2:} How effective is the HTTP hooking approach (in the CEM module) in locating ad traffic from general network traffic?

\noindent
\textbf{RQ3:} How accurate is \tool{} in detecting devious contents?

Our experimental setup includes the construction of a large set of ad-supported apps from markets, the execution of these apps on a physical device, and the collection of network traffic. 

\subsection{Dataset Construction}

To prepare the dataset for evaluating \tool{}, we resort to the well-known AndroZoo dataset~\cite{li2017androzoo++} to crawl Android apps.
Since we are only interested in apps displaying advertisements, we further leverage VirusTotal to collect ad involved apps, i.e., adware.
In this work, we consider a given Android app is adware as long as one anti-virus engine flags it as such.
To this end, we randomly collected 40,000 adware from AndroZoo, including 20,000 Google Play and 20,000 third-party apps, to support our experiment.

Among the randomly selected 40,000 apps, we run each app on a Nexus 5 smartphone, and we use six smartphones in parallel.
Considering that loading an ad from a remote server may take time, we set the transition time in app automation to 5 seconds. Overall, automated exploration for each app takes on average 2 minutes. 
It takes roughly ten days to run all the apps automatically. Contrary to prior related work~\cite{rastogi2016these, chen2019revisiting}, we do not rely on emulators given that ad libraries may implement verification steps to avoid ad networks from serving ads when the app is being experimented on emulated environments~\cite{vidas2014evading} (the objective being to prevent fake impressions of ads~\cite{crussell2014madfraud}, i.e., unjustified profit for app developers).

\subsection{Harvested Ad Content}
\textbf{Ad-related Traffic:}
Out of the 40,000 apps, we were able to successfully run 38,553 (i.e., 96.38\%)  on the Nexus 5 smartphones. 
During the execution of these apps, the TCM module has collected in total 2,488,897 HTTP and HTTPS messages, from which our CEM module flags 541,129 messages related to ad-load (21.7\%) and 692,122 messages related to ad-click (27.8\%). 

\noindent
\textbf{Ad Content:}
The CEM module then extracts ad contents from the collected traffic: we retrieved 83,347 ad images, 52,592 executable scripts, 49,392 redirection URLs, and 2,545 apps directly downloaded and 2,081 apps promoted via Google Play by clicking the ad views.

\subsection{RQ1: Overall Results}

\noindent
\textbf{Devious Content Detection:}
As detailed in Table~\ref{tab:overall_results}, the DDM module flags 279 ad images (specifically, 172 adult, 61 medical, 37 gambling, and 9 violence ad images), 112 executable scripts, 1,822 redirection URLs and 1,457 downloaded apps as devious ad contents. These statistics show that ad clicking contents (i.e., obtained by clicking on displayed ads) are more likely to be devious than ad loading contents (i.e., obtained when loading a page with ad view). 
This is reasonable since dynamic analysis can reveal deviousness if the content is available automatically on the host app (as what ad loading request does) and subsequently may prevent their acceptance on markets. Instead, leaving their loading, at runtime, from third-party servers is a more effective distribution model. 

Nevertheless, although devious ad loading contents are more scarce, they may have a higher impact on the security and privacy of end users.
Indeed, unlike ad clicking contents, which may not be triggered (e.g., the ad is not clicked), ad loading contents will, in any case, be delivered to users when the app is launched.

\begin{table}[!ht]
\centering
\caption{Statistics on harvested ad contents.}
\vspace{-0.1in}
\label{tab:overall_results}
\resizebox{0.8\linewidth}{!}{
\begin{tabular}{l l l l}
\hline
Ad Content                                  & Total & Devious &Type\\ \hline
Ad Images                     & 83,347  & 279 (0.33\%) & Ad Loading\\
Executable Scripts                         & 52,592 & 112 (0.21\%) & Ad Loading  \\
Ad Redirection URLs                      & 49,392 & 1,822 (3.69\%) & Ad Clicking \\
Downloaded Apps                 & 2,545  & 1,457 (57.25\%) & Ad Clicking \\
\hline
\end{tabular}
}
\vspace{-0.1in}
\end{table}

\noindent 
\textbf{Malware Distributed by Ad Content:} It is noteworthy that more than 57\% of the downloaded apps are alerted as suspicious by antivirus engines (hence are categorized as devious content in Table~\ref{tab:overall_results}). More than 30\% of these devious contents are even flagged by over 10 anti-virus engines, indicating a consensus on their maliciousness. Table~\ref{tab:malware} lists the top-3 identified malware ranked by the number of VirusTotal anti-virus engines that agree on them being malicious. We further resort to Google Play and an ASO website (www.chandashi.com) that contains apps in more than 10 third-party markets to search for these apps (based on their unique identifiers).
Expectedly, 21\% of them (311/1457) are not hosted on any markets (both official and alternative markets), and over 91\% (1332/1457) of them are not listed on Google Play. This result suggests that attackers are leveraging ad contents as a new channel to distribute malicious apps, especially considering that more and more app markets are enforcing strict security checks.

\begin{table}[t!]
\newcommand{\tabincell}[2]{\begin{tabular}{@{}#1@{}}#2\end{tabular}}

\centering
\caption{Top 3 downloaded malicious apps ranked by the number of flagged VT engines.}
\vspace{-0.1in}
\label{tab:malware}
\resizebox{\linewidth}{!}{
\begin{tabular}{r l l c}
\hline
Package Name  & MD5 & Source App & \# Engines \\ \hline

com.zhulai.jingjimoren & d3a6fa8359ad1b139004e617ce3baab8 & com.ziipin.softkeyboard.kazakh & 44 \\
\hline
girl.game.weaimeng & 21846ecdfe3ae93391372bdd1cd43032 & air.com.aoaogame.game34  &  37\\
\hline
afltr.austscf798.zhnf760 & e260fd6f711aea632bb4aae2776a1cef & com.easaa.c000000021 & 31 \\
\hline
\end{tabular}
}
\vspace{-0.1in}
\end{table}

\noindent
\textbf{Host Apps of Devious Content:}
We further look into the host apps of these devious content to investigate the spread of devious ad contents. Results are summarized in Table~\ref{tab:hostapps}. Roughly 6.02\% of apps (2,322 out of 38,553) in our dataset are identified as delivering devious ad contents.
The fact that more devious contents are collected than the number of host apps suggests that one app may repeatedly present devious ad contents. Sometimes the same app may present a diversity of devious ad contents.
Moreover, even popular apps on the official Google Play market (e.g., the popular ``Magic Candy'' game app\footnote{By the time of this study, this app is still available on the Google Play market~\cite{MagicCandy} and has received more than 10 million installs.}) are involved in providing devious ad contents (distributing click deceptive images). This is evidence that the identification and blocking of devious ad content remains an unresolved issue in the industry. The research community thus needs to put more effort into approaches and tools for addressing unethical behavior in mobile ads so as to provide a clean and safe environment for displaying mobile ads. \\

\vspace{-2mm}
\begin{table}[h!]
\centering
\caption{Host apps of devious content.}
\vspace{-0.1in}
\resizebox{0.8\linewidth}{!}{
\label{tab:hostapps}

\begin{tabular}{r c c}
\hline
Type & \# Devious Contents & \# Host Apps \\ \hline
Click-deceptive Image     & 525       & 40                  \\ 
Censored Image			  & 279		 & 240 \\
Malicious Script            & 112       & 46                  \\ 
Malicious Redirection Link & 1,822     & 838                \\ 
Malicious App    			 &  1,457     &  1,267                \\
Total                       &  -    & 2,322                \\ \hline
\end{tabular}
}
\vspace{-2mm}
\end{table}

\noindent
\textbf{The Role of Ad Networks:}
We further investigate the distribution of ad networks in terms of the number of devious ad contents that they push to app users' devices.
In this work, we have identified in total 3,518 ad host names (or networks in simplicity).
Due to space limitation, we only listed the top 3 ad networks that distribute devious ad content for each group, as shown in Table~\ref{tab:adnetwork}.
It is interesting to observe that, for censored images, malicious scripts and malicious links, most of them are distributed by popular ad networks. For example, over 46\% of malicious links were distributed by startapp and the google ad network. 
Considering that these popular ad networks are widely adopted, many users may have already been affected by the presence of devious ad contents on their devices. For deceptive images and malicious apps, most of them were found in less-popular ad networks.
We argue that the ad networks need to be responsible for such threats by implementing adequate means to keep devious ad contents from being pushed to end users.

\begin{table}[t]
\small
\centering
\caption{Top ad networks that distribute devious ad content.}
\vspace{-0.1in}
\resizebox{\linewidth}{!}{

\begin{tabular}{|l|l|l|l|}
\hline
\multicolumn{3}{|c|}{\textbf{Top 3 ad networks ranked by the number of distributed click deceptive images}}                      \\ \hline
\textbf{ad network}         & \textbf{\# devious content}              & \textbf{\% devious content}\\ \hline

me2s.co &  382 & 72.8\% \\
go2s.co &  120 & 22.9\% \\
droidhen.com &  8 & 1.5\% \\
\hline
\hline

\multicolumn{3}{|c|}{\textbf{Top 3 ad networks ranked by the number of distributed censored images}}                      \\ \hline

startappexchange.com & 146 & 52.3\% \\
googleads.g.doubleclick.net &  34 & 12.2\% \\
adeco.com &  16 & 5.7\% \\
\hline
\hline

\multicolumn{3}{|c|}{\textbf{Top 3 ad networks ranked by the number of distributed malicious scripts}}                      \\ \hline

googleads.g.doubleclick.net & 74 & 66.1\% \\
startappexchange.com & 18  & 16.1\% \\
nads.wuaiso.com & 3 & 2.7\%\\
\hline
\hline

\multicolumn{3}{|c|}{\textbf{Top 3 ad networks ranked by the number of distributed malicious links}}                      \\ \hline
startappexchange.com & 496 & 30.1\% \\
googleads.g.doubleclick.net & 267 & 16.2\% \\
mobincube.com & 155 & 9.4\% \\
\hline
\hline

\multicolumn{3}{|c|}{\textbf{Top 3 ad networks ranked by the number of distributed malware}}                      \\ \hline

ie2o.com & 343 & 23.5\% \\
gamezi.com & 217 & 14.9\% \\
td68x.com & 132 & 9.2\% \\
\hline
\end{tabular}
}
\vspace{-0.1in}
\label{tab:adnetwork}
\end{table}

\noindent \textbf{The Origin of Devious Contents:}
We further seek to investigate the advertisers that distribute the devious contents by analyzing the landing page of malicious redirection links and the downloading address of malicious apps. Table~\ref{tab:devious-origin} lists the top 5 for each of them. We observe that most of the malicious links and malware were originated from several specific domains. Top 5 domains occupied over 23\% of malicious links, and over 56\% of malware downloading URLs. This result suggested that some advertisers have the tendency to release malicious contents. Therefore, it is important and urgent to identify them and remove them from all the ad networks.

\begin{table}[t]
\small
\centering
\caption{The origin of devious contents.}
\vspace{-0.1in}
\begin{tabular}{|l|l|l|}
\hline
\multicolumn{3}{|c|}{\textbf{Top 5 landing page domains of malicious redirection links}}                      \\ \hline
\textbf{domain} & \textbf{\# malicious link}              & \textbf{\% malicious link}\\ \hline

revcontent.com  & 145 & 8.0\% \\
take-your-prize-now1.life  & 119 & 6.5\% \\
ds-club.ru & 75 & 4.1\% \\
wolve.pro & 42 & 2.3\% \\
inhabitny.com & 39 & 2.1\%\\
\hline
\hline

\multicolumn{3}{|c|}{\textbf{Top 5 downloading domains of malicious apps}}                      \\ \hline
\textbf{domain} & \textbf{\# malicious apps}              & \textbf{\% malicious apps}\\ \hline

ie2o.com  & 343 & 23.5\% \\
gamezi.com  & 217 & 14.9\% \\
td68x.com & 132 & 9.2\% \\
clouddn.com & 78 & 5.4\% \\
cmbst.cn & 54 & 3.7\% \\
\hline

\end{tabular}
\vspace{-0.1in}
\label{tab:devious-origin}
\end{table}

\noindent \textbf{Comparison with the state-of-the-art:} A recent closest study ~\cite{chen2019revisiting} characterizes the malicious behavior of mobile ad landing pages.
Unfortunately, their tool and dataset are not publicly available. Thus, we can neither apply their approach to the apps we randomly selected in this work nor apply \tool{} to their apps.
Hence, we explain why our approach can collect much more ad contents than theirs according to the design. 
First, like all the other previous studies, Chen \textit{et al.} only focuses on ad clicking contents, letting ad loading contents untouched. 
Second, they only take into account WebView widgets for inferring advertisements. However, we experimentally find that WebView is only used by roughly two-thirds of the advertisements (67.58\%). \emph{ImageView and ViewFlipper widgets have also been frequently leveraged to display ads.}
Third, they exclude all apps with multiple WebViews from their dataset. Our experiment reveals that around 30\% (2,581/8,604) of the apps are involved in two or more distinct ad widgets, while 67.2\% (1,734/2,715) of them contains two or more Webviews.
Finally, the list of ad hosts considered by Chen \textit{et al.} also limits their capability of identifying all advertisements.
As shown in the next section, our HTTP hooking approach can significantly increase the list of ad hosts for identifying advertisements.

\subsection{RQ2: Effectiveness of the HTTP hooking}

The CEM is a key module where non-ad traffic is filtered out through a HTTP hooking approach.
Given that this step is essential to locate and extract ad contents, it is important to assess its effectiveness so as to validate this step in the workflow. 
Recall that the HTTP hooking approach takes as input a set of ad libraries and/or ad hosts and the library-host mapping is built iteratively.
We evaluate our approach through the following three settings.
\begin{itemize}

\item \textbf{S1: Ad libraries only.}
We send only ad libraries (with the ad host set as empty) to evaluate the effectiveness of the HTTP hooking approach. Specifically, 52 popular ad networks, maintained by LibRadar~\cite{ma2016libradar}, are considered.

\item \textbf{S2: Ad hosts only.}
Instead of giving ad libraries as input, we send only ad hosts to run the experiment. Specifically, the 1,315 ad hosts\footnote{This list is continuously being updated. At the time of the study presented by Chen \textit{et al.}~\cite{chen2019revisiting}, the number is 1,183.} leveraged by Chen \textit{et al.}~\cite{chen2019revisiting} in their mobile advertising threats study are considered in this experiment.

\item \textbf{S3: Ad libraries and hosts.}
Finally, we take into account both the aforementioned 52 ad libraries and the 1,315 ad hosts as input to conduct the evaluation.

\end{itemize}

\begin{figure}[t]
	\centering
	\includegraphics[width=0.6\linewidth]{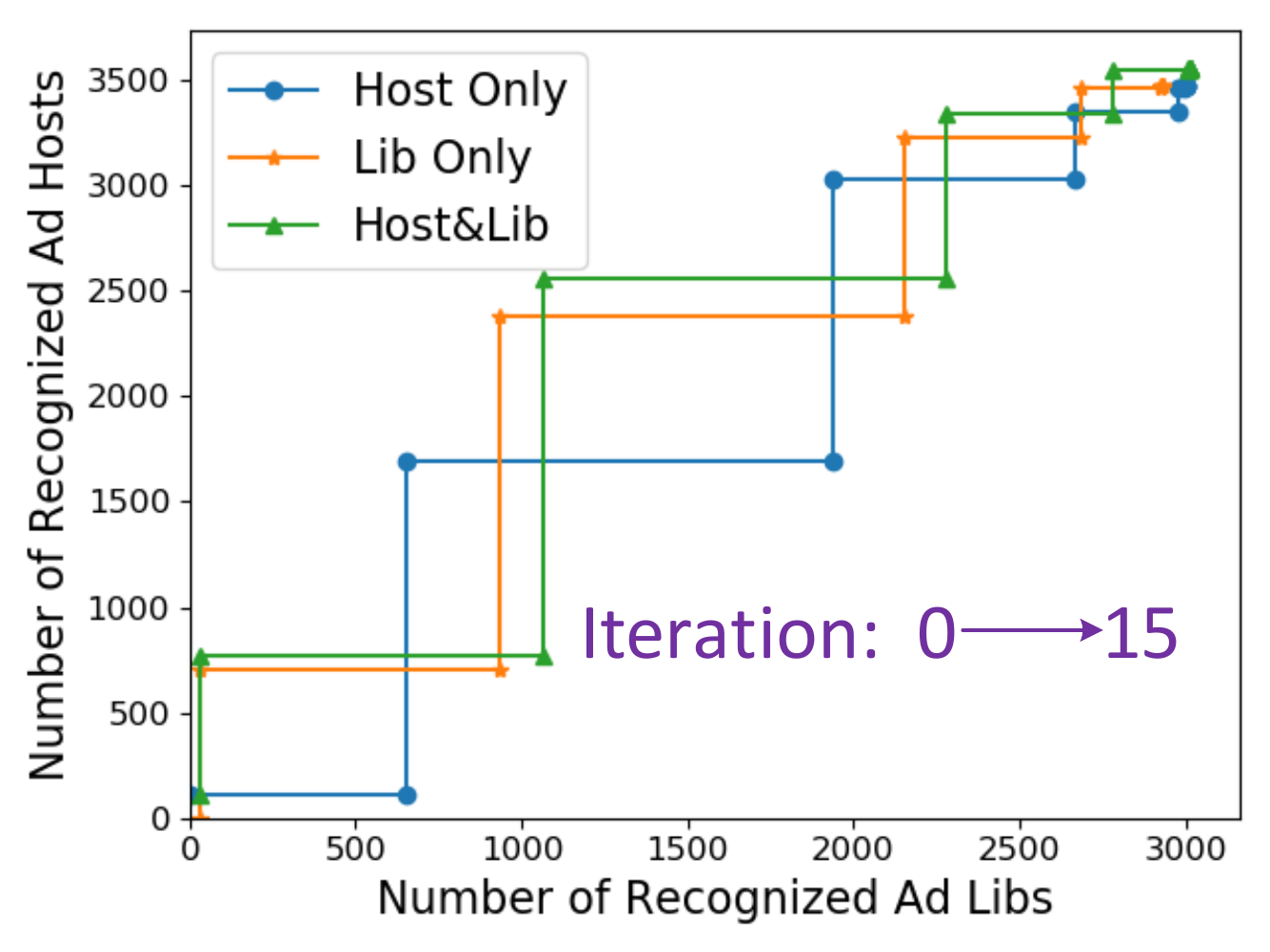}
	\vspace{-0.15in}
	\caption{Results of RQ2 in 15 Iterations. Each iteration is denoted by an edge: vertical edge for expanding ad hosts (based on the latest ad library list) while horizontal edge for expanding ad libraries (based on the latest ad host list).}
    \label{fig:rq2}
    	\vspace{-0.2in}
\end{figure}

Figure~\ref{fig:rq2} illustrates the experimental results.
The \emph{x-axis} and \emph{y-axis} represent respectively the number of ad libraries and ad hosts.
Interestingly, no matter starting from which setting, all the experiments tend to converge to the same result within 15 iterations.
Thanks to the HTTP hooking approach, the number of ad hosts has almost tripled, resulting in around 3,500 ad hosts,
which in turn immensely increases the collection of ad contents by 126\%.

\begin{figure}[t]
    \centering
    \vspace{-0.1in}
    \subfigure[Lib.]{\label{fig:loc}\includegraphics[width=0.49\linewidth]{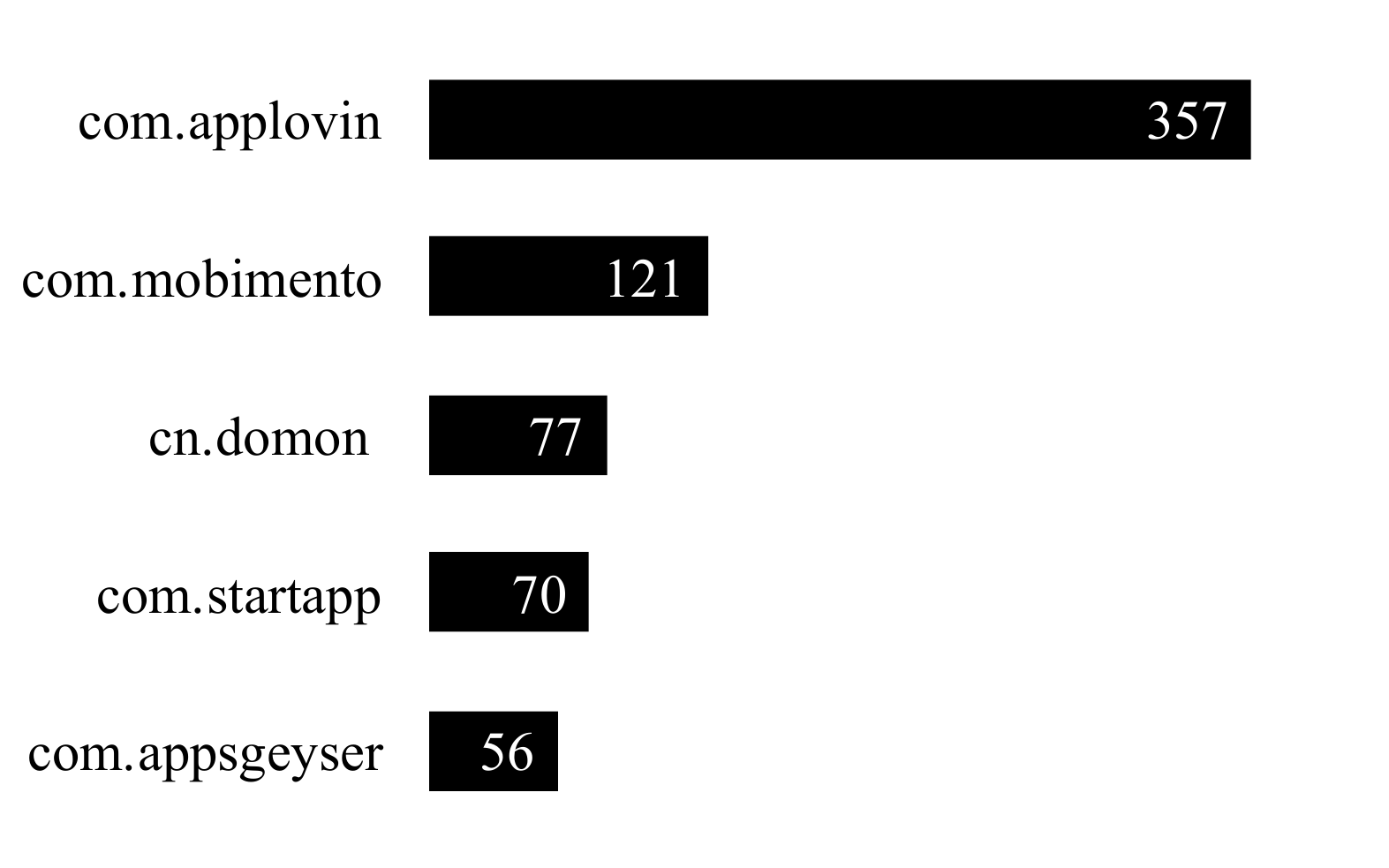}}
    \subfigure[Host.]{\label{fig:cell}\includegraphics[width=0.49\linewidth]{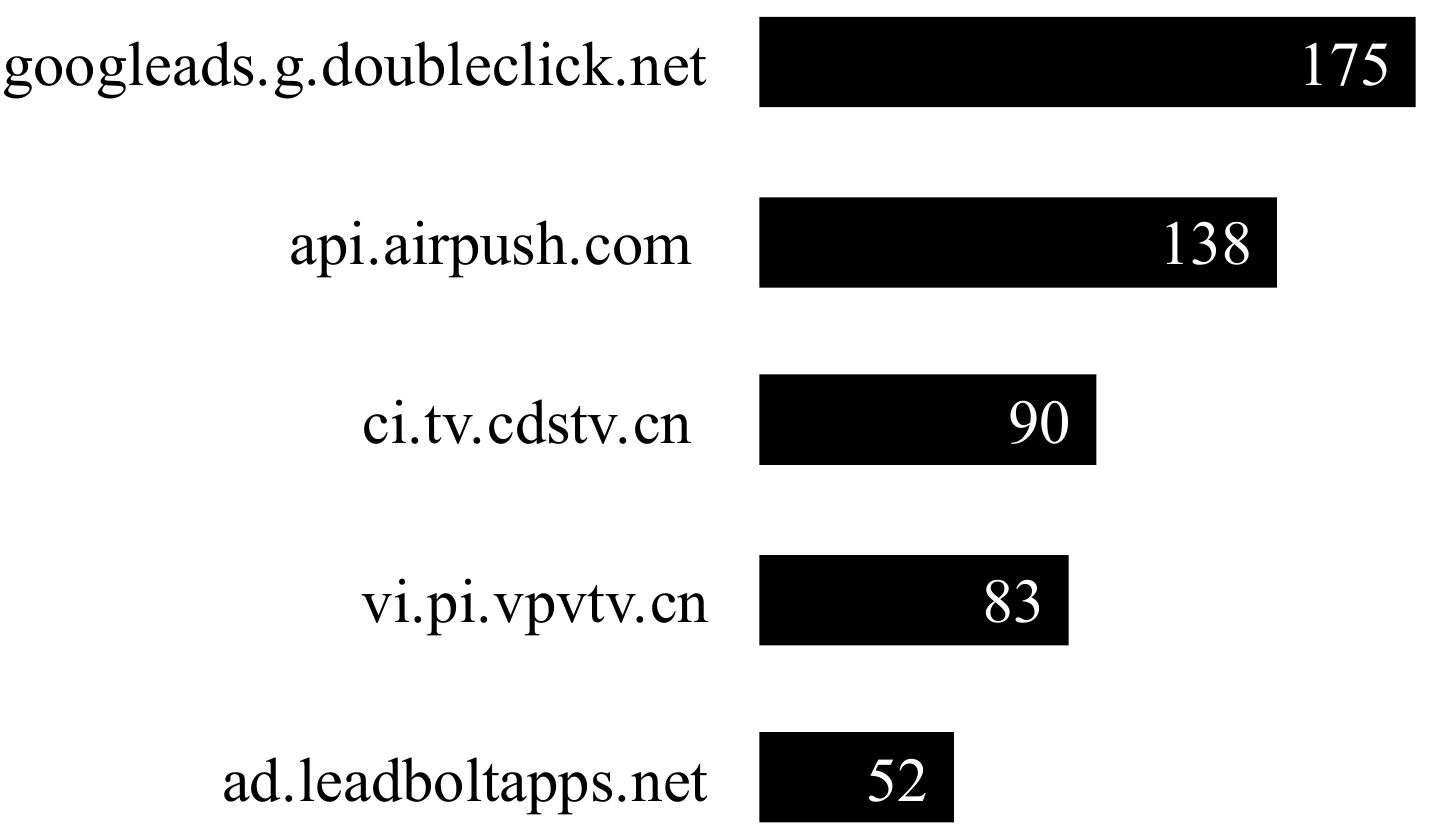}}
        \vspace{-0.15in}
	\caption{Top five involved ad libraries and ad hosts.}
	        \vspace{-0.1in}
    \label{fig:top5}
\end{figure}

Figure~\ref{fig:top5} presents the top five involved ad libraries and ad hosts, w.r.t.  the number of hosts triggered by each library and the number of libraries triggering the same host, respectively.
The top-ranked library, namely \emph{com.applovin}, is even associated with 357 distinct ad hosts, while the top-ranked ad host, namely \emph{googleads.g.\\doubleclick.net}, is triggered by 175 ad libraries.
It is surprising that one ad library can trigger multiple distinct ad hosts and one ad host can be triggered by multiple ad libraries.
Our in-depth manual investigation reveals that those top-ranked libraries have usually embedded with multiple other ad libraries and the actual ad requests are triggered by those embedded ones, resulting in hence different ad hosts mapping to different ad libraries.

Our in-depth analysis further reveals that some libraries pointed to the same ad hosts are actually the same ones but have been deeply obfuscated.
For example, \emph{api.airpush.com} is a well-known ad library. 
In this experiment, we find various libraries such as \emph{com.avtqk.ubjir220086} and \emph{com.filGh.hXwrF124710} that trigger the same ad host (i.e., \emph{api.airpush.com}) and these libraries are actually the obfuscated versions of the original \emph{airpush} library.
This result suggests that our HTTP hooking approach can be even a promising approach for identifying obfuscated libraries.

\subsection{RQ3: Performance Evaluation}

\subsubsection{Click-deceptive Image}
Given that we have built ourselves the training dataset for detecting click-deceptive ad images, and selected a technique that has not been applied in the literature for such cases, we evaluate the performance of this detection as the main pain point in the validation of the DDM module. 
We recall that we have collected 2,375 click-deceptive images as introduced in Section~\ref{sec:ADD}.
To evaluate the performance of our plugin for detecting click-deceptive images, we randomly select 2,175 images to form a training set for YOLOv3.
Then, we put the remaining 200 click-deceptive images, along with 200 normal ad images (i.e., without ``cross'' button embedded) into a testing set and apply the trained YOLOv3 model to it.
Among the 400 images, our YOLOv3 model flags 201 and 199 images as click-deceptive and normal, respectively.
With 5 false positives and 4 false negatives, our approach yields a precision and recall of 97.51\% and 97.99\% respectively for predicting click-deceptive images, demonstrating that our approach is quite reliable for recognizing the ``cross'' button embedded in images.

Most of the closing buttons in ad images follows the form of a cross symbol, but exceptions exist that some of the closing buttons are demonstrated in text images. To this end, we implement a keyword matching based approach to identify \emph{close}, \emph{exit}, and \emph{skip} in Chinese and English. Eventually we identify one such case.

\subsubsection{Censored Images}
Scenarios for detecting Pornographic, Violence, and Medical images are directly based on the popular Google Vision API with pre-trained deep learning models.
Google Vision API has been widely used by state-of-the-art approaches and has been experimentally demonstrated to be effective in flagging pornographic, violence, and medical images~\cite{chen2019revisiting,chen2017content,mazloom2016multimodal,dodge2017parsing}. 
As experimentally demonstrated by Chen \textit{et al.}~\cite{chen2019revisiting}, Google Vision API can indeed outperform other image scanning services. 

For the gambling image detection, with a set of 100 gambling-related keywords in both Chinese and English (configurable), MadDroid achieves 100\% of accuracy for identifying gambling images, as all the identified images indeed contain the defined keywords. Despite that we have formed a relatively large set of gambling-related keywords, it is still possible that some gambling images are overlooked, e.g., they do not contain keywords, or our list of keywords is incomplete. However, these cases are rare in our study.

\subsubsection{Malicious Scripts/Links/Malware}

Since we are not capable of manually confirming if a given app, redirection link or script is malicious, we rely on VirusTotal to flag malicious ones, which is widely used in our research community.

\section{Discussions}

\textbf{Implications.}
Our findings in this paper suggest that ad networks, even popular ones such as AdMob, are involved in the delivery of devious ad contents to end users' devices. The fact that ad networks are not always delivering legitimate ad contents suggests that the ad contents (likely provided by Advertisers) might not be properly checked by ad networks before being pushed to user devices.
As a result, devious Advertisers may exploit the limitations in the current system to advertise devious contents, leading to a poor user experience which harms the reputation of both ad networks and app developers.
We argue that ad networks need to introduce automated tools to regulate the behavior of advertisers. 
Moreover, it is hard to know if ad networks are involved in this black market.
They may turn a blind eye on purpose as they have actually hosted the content servers.
We hence appeal to the community for putting more effort to explore this new research direction.

\textbf{Limitations.}
The implementation of \tool{}, however, carries several limitations. First, we take advantage of state-of-the-art app automation technique~\cite{li2017droidbot} to explore the app, and use an ad-first exploration strategy, to achieve a balance between time efficiency and ad view coverage, which may cause some ad views to be missed during UI exploration. Nevertheless, our experiments suggest that we could extract much more ad contents than existing studies. 
Second, our categorization of devious content might be incomplete since it was built based on existing knowledge. Nonetheless, for new types of devious content, it is quite easy to extend the framework of \tool{} for further analysis. Third, the ad contents shown in a given app may vary due to factors such as time, location, user identifiers, etc. Thus, in our automation testing, some devious behaviors may not be triggered due to various reasons.

\section{Related Work}

\textbf{Mobile Ad Clicking Content Analysis.} Beside the state-of-the-art work by Chen \textit{et al.}~\cite{chen2019revisiting}, the closest work related to ours is proposed by Rastogi \textit{et al.}~\cite{rastogi2016these}, who have experimentally explored the security issues of ad clicking contents, \emph{without considering the ad loading content}.
Similarly, Son \textit{et al.}~\cite{son2016mobile} are also interested in the devious behavior of advertisers.
They have discovered that malicious advertisers may push executable scripts to access the external storage of user's devices so as to infer sensitive information of users.
Our work has also taken into account the aforementioned three kinds of contents, namely malicious link, malware and malicious script. Moreover, to the best of our knowledge, it is the first work that considered the click-deceptive images and censored contents during the loading of mobile ads.

\textbf{Malicious Web Advertising Analysis.}
Malicious advertisement has been extensively studied in the context of web advertising, which is so-called web malvertising. This line of studies mainly falls in the group of drive-by-download attack detection.
Cova \textit{et al.}~\cite{Cova} and Lu \textit{et al.}~\cite{LuBlade} proposed to detect drive-by-download attack and malicious Javascripts that embedded in the advertising.
s. Stringhini \textit{et al.}~\cite{Stringhini} and Mekky \textit{et al.}~\cite{Mekky} used the properties of HTTP redirections to identify
malicious advertisement behaviour.
Li \textit{et al.}~\cite{knowingLi} performed a large-scale study through analyzing ad-related Web traces, and found that malicious advertising infects both top Web sites and leading ad networks (e.g., DoubleClick).

\textbf{Mobile Ad Fraud Detection.}
Various research studies are proposed to investigate the malicious and fraudulent behaviors of mobile app developers~\cite{wang2017explorative, wang2019characterizing}, who aim to entice users to click ads or push notifications~\cite{dong2018frauddroid, liu2014decaf, dong2018mobile, crussell2014madfraud, liu2019dapanda}.
Our approach, targeting the devious behavior of advertisers, can be considered as a supplement of these studies towards building a trustworthy and clean ecosystem for mobile advertising.

\textbf{Mobile Ad Library Detection and Analysis.} The majority of research studies targeting the mobile ad ecosystem are actually focused on ad libraries~\cite{wang2017understanding}.
One line of work focuses on identifying ad libraries~\cite{li2016investigation, ma2016libradar, wang2015wukong, li2017libd}.
The other line of work focuses on the security and privacy issues of ad libraries~\cite{derr2017keep,derr:ccs16,grace2012unsafe,pearce2012addroid, liu2016identifying, li2019revisiting, li2015iccta, zhang2018re}.
Since ad libraries are normally provided by ad networks, who play an important role in distributing ad contents, the aforementioned approaches could be useful for complementing our approach towards better understanding the lifecycle of devious ad contents.

\section{Conclusion}

In this paper, we perform a large-scale characterization study of mobile ad content, which has been largely overlooked by the research community. We first create a comprehensive categorization of devious mobile ad contents, then we build \tool{}, a framework for automated detection of devious mobile ad contents. By applying \tool{} to 40,000 Android apps, we find that devious ad contents are prevalent: 6\% of apps in our study are identified as delivering devious ad contents. To the best of our knowledge, \tool{} is the
first attempt towards mitigating threats from both ad-load and ad-click introduced by mobile ad contents.

\section*{Acknowledgment}
This work was partly supported by the National Natural Science Foundation of China (No.61702045 and No.61772042), by the Hong Kong RGC Projects (No.152223/17E, CityU C1008-16G), by the Australian Research Council (ARC) under projects DE200100016 and DP200100020, 
by the Fonds National de la Recherche (FNR), Luxembourg, under project CHARACTERIZE C17/IS/11693861, by the SPARTA project which has received funding from the European Union's Horizon 2020 research and innovation program under grant agreement No 830892.

\balance

\bibliographystyle{plain}
\bibliography{cite}

\end{document}